\documentclass[12pt]{article}
\usepackage{graphicx}
\begin{document}
\bigskip
\centerline{\Large \bf Should retirement age be coupled to life expectancy ?}

\bigskip
J.S. S\'a Martins$^1$ and D. Stauffer$^2$.

\bigskip
Laboratoire PMMH, Ecole Sup\'erieure de Physique et Chimie
Industrielle, 10 rue Vauquelin, F-75231 Paris, Euroland

\medskip

\noindent
$^1$ Visiting from Instituto de F\'{\i}sica, Universidade
Federal Fluminense; Av. Litor\^{a}nea s/n, Boa Viagem,
Niter\'{o}i 24210-340, RJ, Brazil; jssm@if.uff.br

\noindent
$^2$ Visiting from Institute for Theoretical Physics, Cologne
University, D-50923 K\"oln, Euroland;
stauffer@thp.uni-koeln.de

\bigskip

Abstract: 
Increasing every year the retirement age by the same amount as the increase
of the life expectancy gives roughly stable ratios of the number of 
retired to working-age people in industrialized countries. Continuous influx
of immigrants, below one percent per year of the total population, is needed
for this stabilization.

\bigskip
The increase of the life expectancy (at birth) over the last centuries in the
industrialized countries is
enormous, as seen in Figure 1 from Wilmoth's Berkeley Mortality Database
for Swedish women. The nonlinearity of this increase warns us against 
extrapolating present trends to more than a century. Nevertheless, the 
increase of the number of older people is for the next few decades a rather
predictable ``age quake'' and causes governments in the industrialized
countries to plan reductions in pensions and increases in the retirement age.
In France, these plans lead to social unrest in 2003, in Germany they are
also discussed controversially since fall of 2002, in Brazil the legislation 
of 2003 was pushed through against protest demonstrations of tens of thousands 
of public employees, in California public 
retirement benefits might in the future become available only after an age
coupled to life expectancy. We prefer to simulate the effects
of such legislation on a computer {\bf before} they are imposed on millions
of people (Bomsdorf 1993 and 2003, Tuljapurkar et al. 2000, Olshansky et al.
2001, {\L}aszkiewicz et al. 2003). 

\begin{figure}[hbt]
\begin{center}
\includegraphics[angle=-90,scale=0.5]{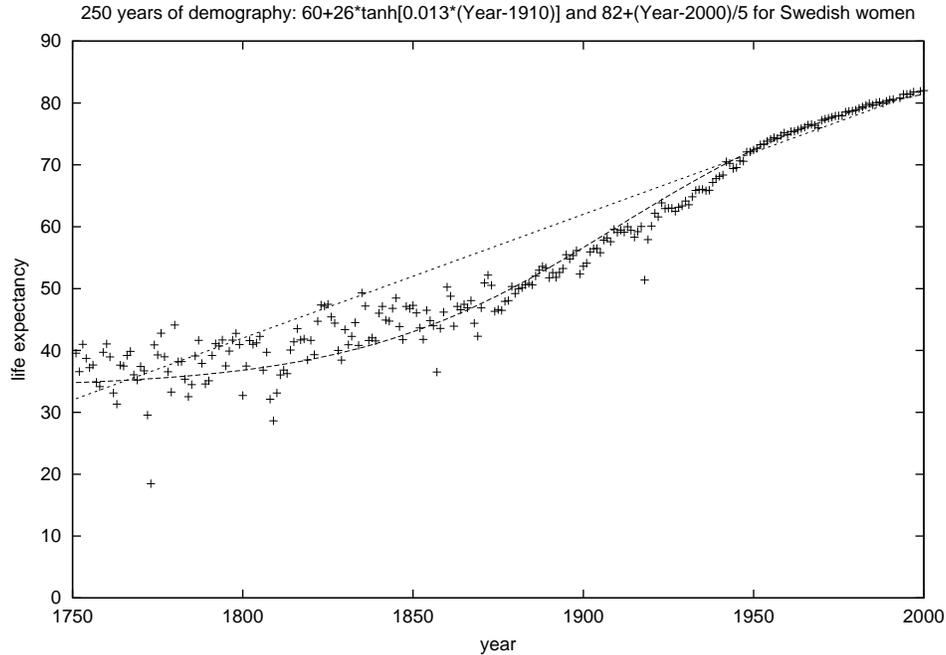}
\end{center}
\caption{
Life expectancy of Swedish women, mostly from 
demog.berkeley.edu/wilmoth/mortality. The approximation by a straight line is
much worse than that by a hyperbolic tangent. Thus our approximations which
lead roughly to straight lines can only be used for limited times.
}
\end{figure}

\begin{figure}[hbt]
\begin{center}
\includegraphics[angle=-90,scale=0.5]{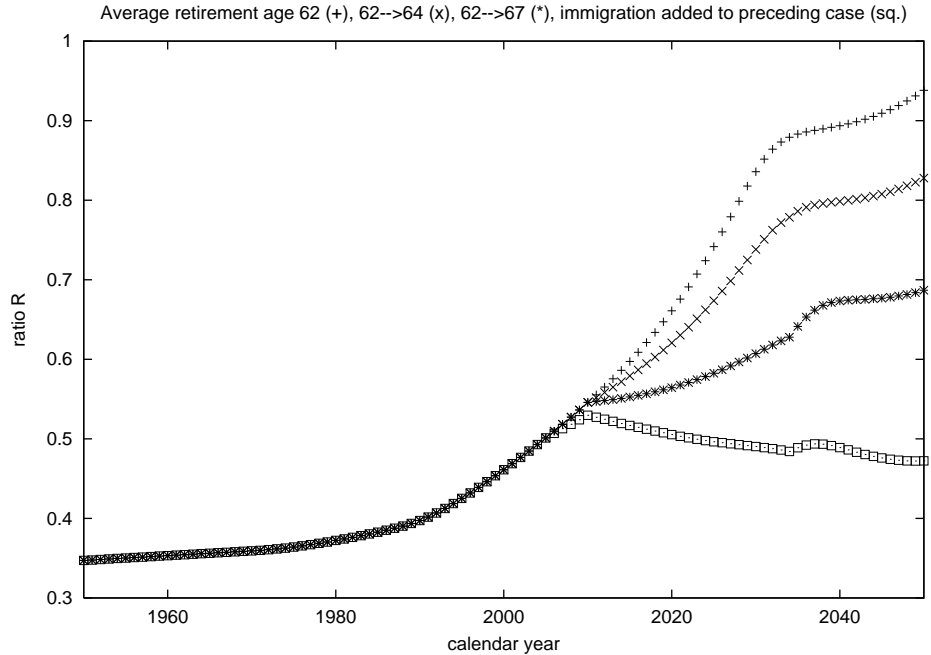}
\end{center}
\caption{
Ratio of retired to working-age people with retirement age changed by 
politicians.
}
\end{figure}

\begin{figure}[hbt]
\begin{center}
\includegraphics[angle=-90,scale=0.5]{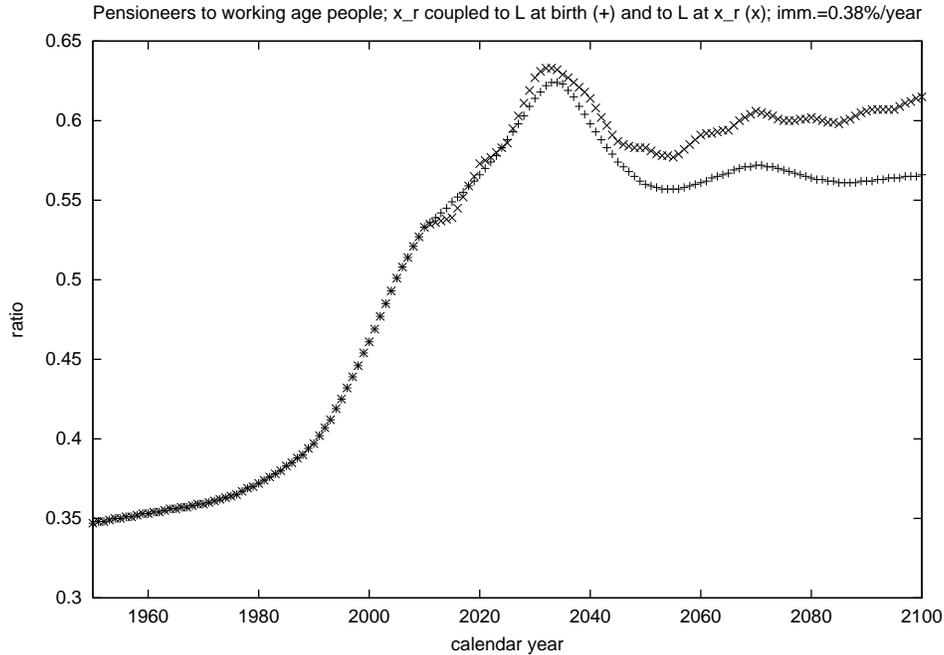}
\end{center}
\caption{
Ratio of retired to working-age people with retirement age changed by changes
in life expectancy.
}
\end{figure}

\begin{figure}[hbt]
\begin{center}
\includegraphics[angle=-90,scale=0.5]{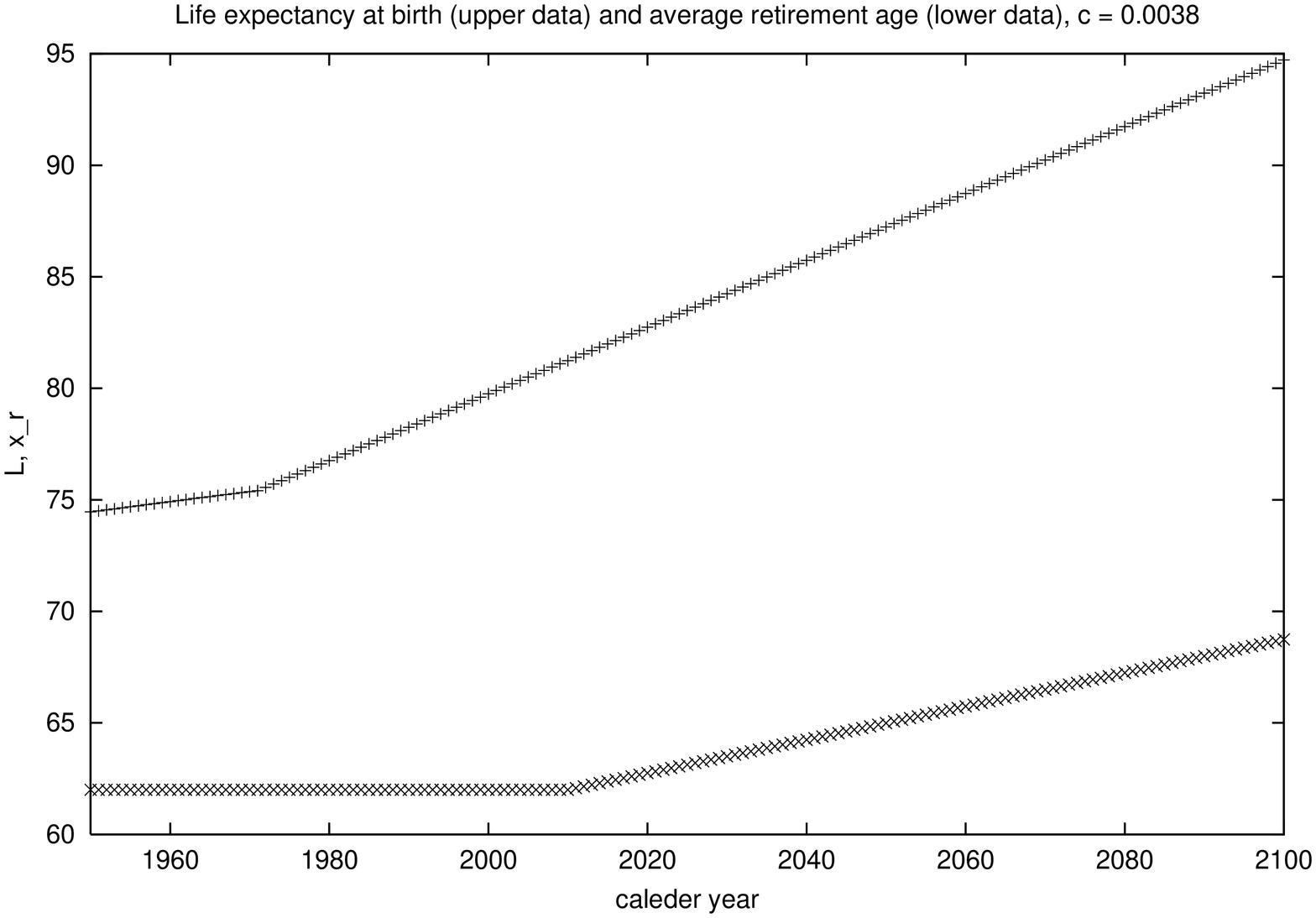}
\end{center}
\caption{
Life expectancy at birth and retirement age for the model of Fig. 3.
}
\end{figure}

Our basic method was described by Stauffer (2002) and assumes for the adult
mortality function at age $x$ a Gompertz law $\mu/b = A \, e^{b(x-X)}$, with 
time-dependent parameters $b \simeq 0.1; X \simeq 100$, where our time unit is
one year (Mildvan and Strehler 1960, Gavrilov and Gavrilova 1991, Azbel 1996,
Wachter and Finch 1997). 
Starting in the year 2005, immigration of people between the ages
of 6 and 40 amounts to a fraction $c$ of the total population each year; $c$ 
is about half a percent. Births diminished drastically around 1970 to
1.4 per woman and are assumed to stay at this value, below the replacement
value slightly above 2; thus immigration is needed to stabilize the population.
The results are given by Stauffer (2002). 

Now we are interested in the ratio $R$ of the number of people beyond average 
retirement age $x_r$ to the number of working-age people and how $R$ depends on 
changes in this retirement age $x_r$. Working was assumed to start at age 20; 
the present retirement age was taken as 62. Figure 2 summarizes the resulting
ratio $R$ from several assumptions. The three top curves assume no immigration.
The top curve assumes the retirement age to stay at 62. The second-highest 
curve assumes $x_r$ to increase from 62 to 64 over an interval of 24 years,
starting in 2011, while the third curve from top assumes an increase from 62 
to 67 over the same time interval; such laws are presently discussed in 
Germany. The lowest curve adds half a percent immigration per year to the model
given by the third-highest curve. 

Such increases in $x_r$ are easily accepted by computers but not by humans. 
Public acceptance might be higher if the changes do not seem to be arbitrarily
imposed by politicians but arise more unavoidably from nature, like ageing.
The life expectancy $L$ at birth is a widely reported quantity, and a coupling
of  
changes in $L$ to changes in $x_r$ appears more plausible and thus perhaps 
more acceptable.  Though $L$ from cohort life tables should be better (Bomsdorf 
2003) than $L$ from period life tables, we take $L$ as that calculated 
from the mortalities in the given year of the computer simulation, since this 
$L$ is best known to the general public.  And we use both $L$ at birth and
the remaining life expectancy after retirement which is more relevant for 
financing retirement than life expectancy at birth. 

Figure 3 shows what happens if, starting after 2010,  each year $x_r$ is 
increased by an amount proportional to the increase of $L$ five years earlier. 
The proportionality factor is 1.0 for $L$ at retirement and 0.6 for $L$ at 
birth. We see 
slight periodic oscillations not visible in Figure 2, but otherwise the 
results look nice and show the dangerous peak in $R$ around the year 2030
to be of rather limited duration and thus perhaps better manageable than the 
results of Fig.2. (Immigration was set at $c = 0.0038$ to keep the total 
population constant in the second half of the 21st century). Fig. 4 shows
the resulting change in $L$ and $x_r$.

In summary we found in Fig.3 a surprisingly stable though high ratio $R$ of 
pensioneers to working-age people, if the retirement age is coupled to the life 
expectancy at birth.

\medskip
\noindent {\bf Acknowledgements}: To PMMH at ESPCI for the warm hospitality,
to Sorin T\u{a}nase-Nicola for helping us with the computer facilities;
JSSM thanks the Brazilian agency FAPERJ for financial support.
\newpage

\parindent 0pt
\bigskip
Azbel, M. Ya., 1996. Unitary mortality law and species-specific age.
Proc. Roy. Soc. B 263, 1449-1454.
 
Bomsdorf, E., 1993. {\it Generationensterbetafeln f\"ur die
Geburtsjahrg\"ange 1923-1993: Modellrechnungen f\"ur die Bundesrepublik
Deutschland}, Verlag Josef Eul, K\"oln; and 2003 preprint, Faculty of Economic 
and Social Sciences, Cologne University.
 
Gavrilov, L.A., Gavrilova, N.S. 1991. {\it The Biology of Life Span},
Harwood Academic Publisher, Chur and 2001. The reliability theory of aging and
longevity, J. Theor. Biology 213, 527-545.

{\L}aszkiewicz A., Szymczak Sz., S. Cebrat S., 2003. 
The Oldest Old and the Population Heterogeneity. Int. J. Modern Physics C 14,
No. 10 (2003).

Olshansky, S.J., Carnes B.A., D\'esesquelles A., 2001.
Demography - Prospects for human longevity Science 291, 1491-1492.
 
Stauffer D., 2002.  
Simple tools for forecasts of population ageing in developed countries
based on extrapolations of human mortality, fertility and migration.
Exp. Gerontology 37, 1131-1136.

Strehler, B. L., Mildvan, A.S. 1960. General theory of mortality
and aging, Science 132, 14-21.

Wachter, K.W., Finch, C.E., 1997. {\it Between Zeus and the Salmon. The
Biodemography of Longevity}, National Academy Press, Washington DC; see
also La Recherche 322 (various authors), July/August 1999;
Nature 408, No. 680 (various authors), November 9, 2000.
 
Tuljapurkar, S., Li, N., Boe, C., 2000. A universal pattern of mortality
decline in the G7 countries. Nature 405, 789-792.
 
\end{document}